\documentclass[12pt]{article}
\usepackage{amssymb, amsmath, amsfonts, mathrsfs, graphicx, subfigure, epsfig, psfrag, arydshln, multirow,  pdflscape, natbib}

\usepackage{bm}
\usepackage{tikz}
\usetikzlibrary{patterns}
\usepackage{pict2e,picture}
\usepackage[figuresright]{rotating}

\usetikzlibrary{shapes.symbols, shapes.geometric, arrows, positioning, chains}

\newcommand{\bepsilon} {\mbox{\boldmath $\epsilon$}}

\newcommand{\bSigma} {\mbox{\boldmath $\Sigma$}}

\newcommand{\bpsi} {\mbox{\boldmath $\psi$}}
\newcommand{\bPsi} {\mbox{\boldmath $\Psi$}}
\newcommand{\bb} {\mbox{\boldmath $b$}}

\newcommand{\be} {\mbox{\boldmath $e$}}

\newcommand{\bu} {\mbox{\boldmath $u$}}

\newcommand{\bX} {\mathbf {X}}

\newcommand{\bW} {\mathbf {W}}
\newcommand{\bU} {\mathbf {U}}
\newcommand{\bV} {\mathbf {V}}

\newcommand{\bT} {\mathbf {T}}
\newcommand{\bA} {\mathbf {A}}
\newcommand{\bB} {\mathbf {B}}
\newcommand{\bC} {\mathbf {C}}
\newcommand{\bE} {\mathbf {E}}
\newcommand{\bF} {\mathbf {F}}
\newcommand{\bH} {\mathbf {H}}
\newcommand{\bI} {\mathbf {I}}
\newcommand{\bP} {\mathbf {P}}
\newcommand{\bQ} {\mathbf {Q}}
\newcommand{\bR} {\mathbf {R}}
\newcommand{\bzero} {\mbox{\boldmath $0$}}
\newcommand{\bone} {\mbox{\boldmath $1$}}
\def\T{{ \mathrm{\scriptscriptstyle T} }}
\DeclareMathOperator*{\argmin}{arg\,min}

\begin{document}

\title{Corrected score methods for estimating Bayesian networks with error-prone nodes}
\author{{\small Xianzheng Huang$^{*1}$, Hongmei Zhang$^2$}\\
{\small $^1$Department of Statistics, University of South Carolina, Columbia, SC 29208, U.S.A.}\\
{\small $^2$Division of Epidemiology, Biostatistics, and Environmental Health,}\\
{\small University of Memphis, Memphis, TN 38111, U.S.A.}\\
{\small $^*$huang@stat.sc.edu}}
\date{}
\maketitle

\begin{abstract}
Motivated by inferring cellular signaling networks using noisy flow cytometry data, we develop procedures to draw inference for Bayesian networks based on error-prone data. Two methods for inferring causal relationships between nodes in a network are proposed based on penalized estimation methods that account for measurement error and encourage sparsity. We discuss consistency of the proposed network estimators and develop an approach for selecting the tuning parameter in the penalized estimation methods. Empirical studies are carried out to compare the proposed methods and a naive method that ignores measurement error with applications to synthetic data and to single cell flow cytometry data.
\end{abstract}


\section{Introduction}
\subsection{Motivations}
The study of cellular signaling networks has been a major research area in biology for several decades. By analyzing how multiple cell signaling pathways affect each other and, in turn, cellular processes within a network, scientists gain valuable insights on normal cellular responses in a biological system, and their potential disregulation in disease \citep{jordan2000signaling, quaranta2013lies, Madireddy2019}. Statistical models that mathematically conceptualize these signaling networks have been developed, which advance experimental cell biology, and influence the way biologists view, monitor, and study signaling networks by perturbing them in designed experiments \citep{janes2013models, karamouzis2014tackling}. 

Among these models, Bayesian networks \citep{Jensen1996} have been widely adopted as an attractive model for characterizing complex cell signaling cascades. With recent advances in biochemistry, molecular biology, and cell physiology, rich data information become available at the cell level from high throughput technologies. For example, flow cytometry is an important tool in a broad range of biological and clinical research, which makes  measuring  physical and chemical characteristics of cells possible. This technology produces abundant data that can be used to infer cellular signalling networks \citep{Sachs05, friedman2008sparse, Shojaie10, luo2011bayesian, Fu13}. However,  measurement errors in flow cytometry data inevitably arise from imperfect measurements, photon-counting statistics, and data storage methods \citep{roederer2001spectral, petrunkina2010systematic, tiberi2018bayesian, galbusera2019using}. This motivates our study presented in this article, where we develop methods for inferring Bayesian networks representing cellular signaling networks using error-prone flow cytometry data. 

The proposed methods can be used to infer Bayesian networks arising in other applications, such as for constructing social networks based on survey data subject to imperfect respondent recall \citep{wang2012measurement}, studying connectivity and association between different regions of one's brain in the default mode network using preprocessed noisy brain image data \citep{li2013bayesian}, and modeling gene regulatory pathways using gene expression data that are prone to measurement error due to experimental errors \citep{ma2006data}, stray background signal irrelevant to mRNA transcripts \citep{strimmer2003modeling}, or data normalization \citep{evans2016selecting}. It is thus instructive to have an overview of literature on networks and network inference in a general context next. 

\subsection{Literature Review}
Networks, or graphs, have been a topic of great interest that started mostly in the artificial intelligence community \citep{Jensen1996, Neapolitan2012, Pearl2014}. Later its application became more widespread, motivating statistical research on graphical models used in biology, genetics, social science, and physics \citep{Lauritzen1996, Edwards2012}. A network consists of a set of nodes, also referred to as vertices or variables, and a set of edges connecting nodes. Graphs with undirected edges are called undirected graphs. In an undirected graph, a set of nodes connecting to a particular node form a neighborhood of this node. Given its neighborhood, this node is independent of nodes outside of the neighborhood. This type of graphs is useful for characterizing correlations between nodes. When causal relationships are of interest, directed edges are used, giving rise to the so-called directed acyclic graphs (DAG). Pairing such a graph with a joint probability distribution of all nodes produces a Bayesian network. When there is an edge pointing from one node to another node, these two nodes are referred to as a parent node (of the latter) and a child node (of the former), respectively. Given its parents, a node is independent of its non-descendant nodes, which is more formally known as the local Markov property of DAG. In this sense, a Bayesian network encodes the joint distribution of the set of nodes in the graph. Provided with this encoding, not only can one uncover the correlation structure among nodes, but one can also reveal if a correlation between two nodes is due to a direct causal relationship between them or an indirect dependence mediated by other nodes. The latter piece of information is especially of interest in biology and genetics \citep{friedman2000using}. Because of this, some researchers refer to Bayesian networks as causal networks to signify causality as their research focal point, as in our motivating study of cellular signaling networks.

There is a large collection of works on inferring Bayesian networks. Many existing works follow the theme of search-and-score \citep{suzuki1993construction, heckerman1995learning, xiang1997microscopic, friedman1999learning, chickering2002learning, moore2003optimal, bartlett2017integer, correia2019pruning}. Following this theme, one formulates a scoring criterion, and searches for a directed graph, or an equivalent class of directed graphs \citep{andersson1997characterization}, that optimizes the score. The score can be constructed based on a likelihood function of observed data in the frequentist framework \citep{Shojaie10}; it can also originate from a posterior distribution of a graph in the Bayesian framework \citep{heckerman1995learning}. Scores formulated borrowing these two schools of statistics have also been used, such as the Bayesian Dirichlet equivalent uniform score defined as the log likelihood of the observed data given suitably chosen Dirichlet priors over the parameters of a network structure \citep{correia2019pruning}. When the number of nodes is large, scores designed to penalize complexity of a graph are often employed \citep{alon1995color, van2013ell}. Another well explored theme for inferring Bayesian networks leads to the constraint-based approaches that involve testing conditional independence among nodes \citep{spirtes1991algorithm, spirtes2000causation}. To lessen the computational burden in the presence of many nodes, \cite{tsamardinos2006max} proposed the max-min hill-climbing algorithm that combines ideas from search-and-score, constraint-based approaches, and local learning. \cite{friedman2003being} used a Markov chain Monte Carlo (MCMC) method over the space of node orders, which is smaller and more regular than the space of graph structures. \cite{eaton2012bayesian} suggested to apply dynamic programming algorithm on the space of node orders, then used the resultant proposal distribution for MCMC methods in the DAG space. Also considering the order space, \cite{ellis2008learning} developed a fast MCMC algorithm based on data that include interventional data and observational data. \citet{ye2019optimizing} proposed to minimize a regularized Cholesky score over the space of topological orderings and achieved improved performance in network structure learning compared with several competing methods when applied to both observational and interventional data. Interventional data arise from intervention experiments, such as flow cytometry experiments considered in our study. In such an experiment, one forces the values of some node(s) to be certain values, which in effect destroys the causal dependencies of the intervened node(s). Inclusion of interventional data greatly improves the identifiability of a Bayesian network, as \cite{hauser2012characterization} explained in great detail. 

All aforementioned existing works rely on observed data as precise measures of nodes. But, as seen in the motivating examples, measures of nodes can be imprecise. For flow cytometry experiments, \cite{galbusera2019using} showed that flow cytometry measurements contain a significant amount of shot-noise that can be easily mistaken for true biological variability. Although measurement error problems have been long investigated in many regression settings \citep{Carroll06, fuller2009measurement, grace2016statistical}, there is very limited research in the context of inferring Bayesian networks. One exception is the work by \cite{luo2011bayesian}, who used Bayesian hierarchical modeling to incorporate measurement error and random error that represent intrinsic noise in flow cytometry data when inferring signaling pathways. In this article, we tackle this problem from the frequentist point of view. To the best of our knowledge, this is the first frequentist work addressing this problem. 

The data structure considered in our study and mathematical formulations of the data generating mechanism are described in Section~\ref{s:datamodel}. We then outline the proposed penalized estimation methods in Section~\ref{s:estimateB}, which includes detailed algorithms for implementing the proposed methods. To choose the tuning parameter in the penalized estimation, we construct a tuning parameter selector in Section~\ref{s:chooselam}. In Section~\ref{s:simulation}, simulation studies are reported, where we compare finite sample performance of the proposed methods and a naive method that ignores measurement error. We also apply these methods to a flow cytometry data set to infer a signaling network of immune system cells. In Section~\ref{s:discussion}, we summarize the contribution of our study and discuss follow-up research. 

\section{Data and Model}
\label{s:datamodel}
Denote by $\bX$ the $N \times p$ (unobserved) data matrix as error-free measures of $p$ nodes in a network, including interventional data and observational data from $N$ experimental units. Refer to node $j$ as $X_j$, denote by $n_j$ and $n_{-j}$ the number of interventional data points and the number observational data points associated with $X_j$, respectively, and by $O_j$ the set of row indices corresponding to the observational data for $X_j$ in $\bX$, for $j=1, \ldots, p$. The observed data matrix of the same dimension, $\bW$, is an error-contaminated surrogate of $\bX$. 

Taking the data structure into consideration, we assume that the causal relationships of the $p$ nodes are specified by 
\begin{equation}
\bX[O_j, j] = \bX[O_j, -j]\bB_j+\bepsilon[O_j, j], \textrm{ for $j=1, \ldots, p$},\label{eq:DAGx}
\end{equation}
where $\bepsilon$ is the $N\times p$ matrix of model error representing intrinsic noise due to unmodelling variation, $\bepsilon[O_j, j]$ consists of $n_{-j}$ independent and identically distributed (i.i.d.) mean-zero random errors, $\bB=[\beta_{ij}]_{i,j=1, \ldots, p}$ is the $p\times p$ matrix of regression coefficients with zero diagonal entries, and $\bB_j=\bB[-j, j]$. The regression model representation of a Bayesian network in (\ref{eq:DAGx}) is the same as that formulated in \citet{Fu13}. It is assumed that $\bb=(\bB_1^\T, \ldots, \bB_p^\T)^\T$ is a vector of natural parameters in the sense that, given sufficient interventional data associated with each node, $\bb$ is identifiable \citep{Fu13}. For $X_j$, the nodes on the right-hand-side of (\ref{eq:DAGx}) associated with nonzero entries in $\bB_j$ are  parents of $X_j$. Having $\bB_j=\bzero$ means that $X_j$ has no parent, and is referred to as a root node. Having the $j$th row, $\bB[j, ]$, as a zero vector implies that $X_j$ a childless node. Assume that $\bW$ results from contaminating $\bX$ with additive mean-zero normal measurement error independent of $\bX$, that is,
\begin{equation}
\bW=\bX+\bU, \label{eq:wxu}
\end{equation}
where $\bU$ is the $N \times p$ matrix of nondifferential measurement error \citep[][Section 2.5]{Carroll06}. It is further assumed in this study that, for each node $X_j$, the measurement error associated with the interventional data of $X_j$ and the measurement error associated with the observational data of $X_j$ follow the same distribution. This implies that $\{\bU[\ell, \,\,]\}_{\ell=1}^N$ are i.i.d. random vectors from $N_p (\bzero, \bSigma_u)$, where $\bSigma_u$ is the $p \times p$ variance-covariance matrix of the measurement error associated with nodes $(X_1, \ldots, X_p)$. 

According to (\ref{eq:DAGx}) and (\ref{eq:wxu}), the Bayesian network with error-prone nodes consists of $p$ hierarchical measurement error models, with the $j$th hierarchical model consisting of two submodels, 
\begin{eqnarray}
\bW[O_j, j] & = & \bX[O_j, -j]\bB_j+\bepsilon[O_j, j]+\bU[O_j, j], \label{eq:regwx} \\
\bW[O_j, -j] & = & \bX[O_j, -j]+\bU[O_j, -j], \label{eq:wxadd}
\end{eqnarray}  
where the first submodel is for the error-contaminated node $j$ regressing on the remaining $p-1$ error-free nodes, and the second submodel relates the observed covariates with the true covariates in the $j$th regression model, for $j=1, \ldots, p$. Given the set of $p$ measurement error models, making inference for an underlying Bayeisan network that relates the $p$ nodes mainly involves inferring $\bB$ using $\bW$. The variance-covariance associated with $\bepsilon[O_j, j]$ does not need to be estimated for the proposed methods. Estimating $\bSigma_u$ requires either external validation data or replicate measures of the same set of error-free measures of nodes. For instance, it has been a routine practice in the measurement error literature that, with replicate measures on the true covariates, one can use equation (4.3) in \cite{Carroll06} to estimate $\bSigma_u$, which usually has little impact on the final inference on regression parameters. In order to focus on inference on $\bB$, we assume $\bSigma_u$ known in this study. 

\section{Estimation of $\bB$}
\label{s:estimateB}

\subsection{Penalized Objective Functions}
\label{s:obj}
When $\bX$ is observed, \cite{Fu13} proposed to estimate $\bB$ via minimizing a penalized log-likelihood function corresponding to the graphical model in (\ref{eq:DAGx}). In the presence of measurement error, a naive approach for estimating $\bB$ is to ignore measurement error and use $\bW$ in place of $\bX$ in the penalized log-likelihood function in \cite{Fu13}, 
\begin{equation}
R_{\textrm{nv}}(\bB)=\sum_{j=1}^p\left\{V_{j,\textrm{nv}}+\sum_{i=1}^p P_\lambda(|\beta_{ij}|)\right\},
\label{eq:naiveobj}
\end{equation}
where, for $j=1, \ldots, p$, 
\begin{equation}
V_{j,\textrm{nv}}=\frac{n_{-j}}{2}\log\left\{\sum_{\ell\in O_j}\left(\bW[\ell, j]-\bW[\ell, -j]\bB_j\right)^2 \right\},
\label{eq:naiveVj}
\end{equation}
and $P_\lambda(\cdot)$ a penalty function. One may choose a penalty according to the LASSO \citep{Tibs96}, the adaptive LASSO (ALASSO) \citep{Zou06}, or the SCAD penalty \citep{Fan01}. Both ALASSO and SCAD have been shown to enjoy the appealing oracle properties in variable selections. In this study we adopt SCAD in (\ref{eq:naiveobj}), defined as 
\begin{align*}
P_\lambda(t) = & \lambda t I(t \in [0, \lambda))+\displaystyle{\frac{(a^2-1)\lambda^2-(t-a\lambda)^2}{2(a-1)}I(t\in [\lambda, a\lambda))}\\
& +\displaystyle{\frac{(a+1)\lambda^2}{2} I(t\ge a \lambda)},
\end{align*}
in which $\lambda$ is a tuning parameter and $a=3.7$. Besides avoiding the adaptive weights required in ALASSO, our choice of the SCAD penalty is also motivated by findings in \citet{aragam2015concave}, who showed that a concave penalty, such as SCAD, offers improved performance in Bayesian network structure learning when comparing with an $L_1$-based penalty like LASSO. Denote the estimator of $\bB$ by $\hat \bB_{\textrm{nv}}$, as a minimizer of (\ref{eq:naiveobj}) that induces a DAG. 

To account for measurement error in node data, we construct a penalized objective function based on the corrected score function \citep{Nakamura90}. Assuming normal model error and measurement error, the corrected score function associated with the $j$th measurement error model is given by
\begin{align}
\bPsi_j(\bB_j) & = \sum_{\ell \in O_j}\bPsi_{j \ell}(\bB_j) \nonumber \\
& = \sum_{\ell \in O_j}\left\{(\bW[\ell, j]-\bW[\ell, -j]\bB_j)\bW[\ell, -j]^t+\bSigma_u[-j, -j]\bB_j\right\}. \label{eq:sumscorej}
\end{align}
When $\bSigma_u=\bzero$, the summand in (\ref{eq:sumscorej}), $\bPsi_{j \ell}(\bB_j)$, reduces to the score used in the least squared method for estimating the regression coefficients in the $j$th regression model, for $j=1, \ldots, p$. In the presence of measurement error, one can show that $E\{\bPsi_{j \ell}(\bB^*_j)\}=\bzero$ \citep[][Section A.6]{Carroll06}, where $\bB^*_j$ is the truth of $\bB_j$, for $\ell\in O_j$ and $j=1, \ldots, p$. In other words, the corrected score $\bPsi_{j \ell}(\bB_j)$ is an unbiased score that corrects the score used in the least squared method for measurement error. 

For each $j\in \{1, \ldots, p\}$, we follow the construction of quadratic inference functions \citep{qu2000improving} and propose the penalized score-based objective function given by  
\begin{equation}
R(\bB)=\sum_{j=1}^p\left\{V_j+\sum_{i=1}^p P_\lambda(|\beta_{ij}|)\right\},
\label{eq:bigBhat}
\end{equation}
where 
\begin{equation}
V_j=\left\{\frac{1}{n_{-j}} \sum_{\ell\in O_j} \bPsi_{j\ell}(\bB_j) \right\}^t \{\bH_j(\bB_j)\}^{-1} \left\{\frac{1}{n_{-j}} \sum_{\ell\in O_j} \bPsi_{j\ell}(\bB_j) \right\}, \label{eq:nonpenobj}
\end{equation}
in which $\bH_j(\bB_j)=n_{-j}^{-1} \sum_{\ell\in O_j} \bPsi_{j \ell}(\bB_j)\bPsi_{j\ell}(\bB_j)^t$ is a consistent estimator of the variance-covariance matrix of the corrected score, which is ``sandwiched" between the scores to achieve optimality in efficiency of the score-based inference \citep{Hansen82}. A non-naive estimator of $\bB$, denoted by $\hat \bB$, is a minimizer of $R(\bB)$. 

In Appendix A of the Supplementary Materials, we establish the consistency of the estimator as a minimizer of (\ref{eq:bigBhat}) with a fixed $p$ under regularity conditions. The conclusion is summarized in the following theorem. 

\noindent
{\bf Theorem 3.1} Under assumptions (A1)--(A5) in Appendix A, as $n=\min_{1\le j \le p} n_{-j} \to \infty$, if $\sqrt{n}\lambda_n=o_p(1)$, then there exists a local minimizer of $R(\bB)$ defined in (\ref{eq:bigBhat}), denoted by $\hat \bB$, such that $\|\hat\bb-\bb^*\|=O_p(n^{-1/2})$, where $\lambda_n$ is the tuning parameter $\lambda$ in (\ref{eq:bigBhat}) with the added subscript $n$ to signify its dependence on $n$ in the discussion of asymptotics, $\bb^*=(\bB_1^{*\T}, \ldots, \bB_p^{*\T})^\T$, in which $\bB^*_j=\bB^*[-j,j]$, for $j=1, \ldots, p$, and $\bB^*$ is the true value of $\bB$; $\hat \bb$ is similarly defined from $\hat \bB$.

\subsection{Algorithms for Estimating $\bB$}
To find a minimizer of the penalized log-likelihood, \cite{Fu13} developed a pairwise coordinate descent (PCD) algorithm to iteratively update each of the $p(p-1)/2$ pairs, $(\beta_{ij}, \beta_{ji})$, for $i\ne j=1, \ldots, p$, with all other entries of $\bB$ fixed at their values from the preceding iteration. The algorithm is designed to avoid estimates for $\beta_{ij}$and $\beta_{ji}$ to be nonzero simultaneously, since $\beta_{ij}$ and $\beta_{ji}$ both being nonzero is a violation of acyclicity. But PCD cannot avoid other forms of acyclicity violation. To thoroughly check for cycles in an estimated regression coefficients matrix, we implement Kahn's topological sorting algorithm \citep{kahn1962topological} along with PCD. 

\begin{figure}[h]
\centering
\begin{tikzpicture}[
       font = \footnotesize,%\sffamily,
start chain = going below,
    node distance = 7mm and 10mm,
      base/.style = {draw, on chain, align=flush center,  
                     text width=30mm, minimum height=7ex, inner ysep=1mm},
     start/.style = {base, rectangle, rounded corners, fill=gray!20},
       proc/.style = {base, rectangle},
      test/.style = {signal, base,
                     signal to=east and west,
                     text width=40mm, inner xsep=-1ex},
     arrow/.style = {->, draw, thick},
every join/.style = {arrow},
                        ]
%---
\linespread{0.8}

\node (start) [start] {Input $G$};
\node (findroot) [test, join, right=of start] {Are there root nodes in $G$?};
\node (weaklnk) [proc, below=of findroot, pattern=north west lines, pattern color=gray!35] {Remove the weakest edge from $G$};
\node (add) [proc, right=of findroot] {Add root nodes to the tail of $\mathscr{T}$};
\node (outgo) [proc, join, below=of add] {Remove out-going edges from root nodes};
\node (childless) [proc, join, below=of outgo]{Remove childless root nodes from $G$};
\node (empty) [test, below=of weaklnk]{Is $G$ empty?}; 
\node (end) [start, below=of empty] {Output $\mathscr{T}$}; 

\draw [arrow] (findroot.east) node[above right] {Yes} -- (add);
\draw [arrow] (findroot.south) node[below right]{No} -- (weaklnk);
\draw [arrow] (empty.south) node[below right]{Yes}--(end);
\draw [arrow] (empty.west) node[above left]{No} -| ([xshift=-3mm]start.south);
\draw [arrow] (childless)--(empty.east);
\draw [arrow] (weaklnk.west) node[above right]{} -| ([xshift=+3mm]start.south);
\end{tikzpicture}
\caption{\label{f:topo} Kahn's topological sorting algorithm for eliminating cycles in $G$ and finding a topological ordering, $\mathscr{T}$, compatible with the resultant acyclic graph.}
\end{figure}
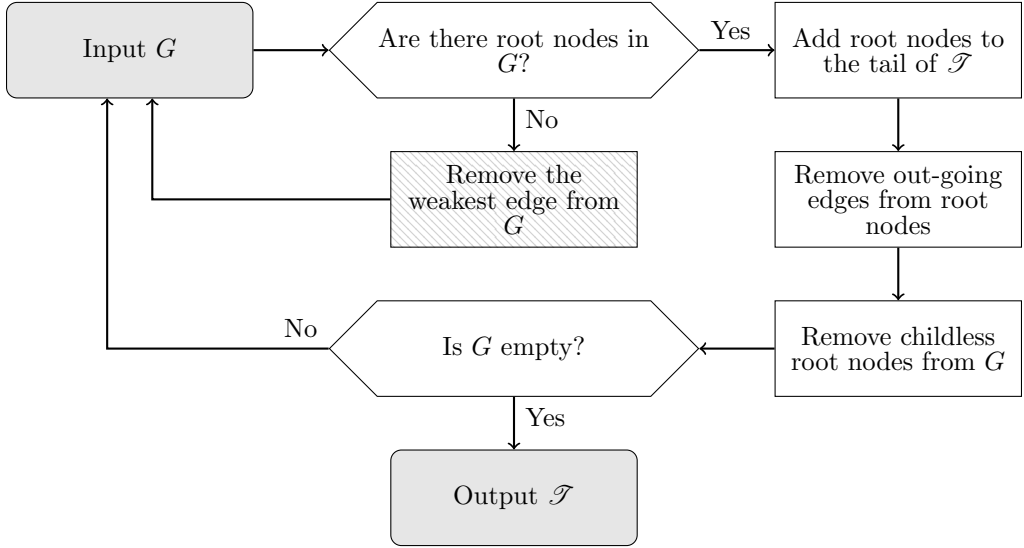

Given a directed graph structure $G$, a topological sorting algorithm is an iterative procedure that yields a sorted sequence of nodes such that a child node always comes after its parent nodes, thus provides an order of these nodes compatible with $G$. A topological sorting algorithm can be used to detect cycles because a topological ordering of nodes does not exist as long as there exists a cycle in the graph \citep{cormen2001introduction}. In particular, Kahn's sorting algorithm is developed based on the fact that a DAG must have at least one root node; moreover, removing root nodes and their out-going edges from a DAG always yields a subgraph that is still a DAG. Hence, an early termination of the sorting algorithm will only occur if a subgraph at that step has no root node, which directly indicates existence of at least one cycle in the subgraph, and thus in the original graph as well. When this occurs, we will strategically remove edges until root nodes emerge so that the sorting algorithm can resume. Figure~\ref{f:topo} illustrates the application of Kahn's sorting algorithm for the purpose of cycle detection and elimination for an initial graph structure as the input of the algorithm. The output of the depicted algorithm is an order compatible with the resultant acyclic graph indicated by a queue of $p$ nodes, denoted by $\mathscr{T}$, which starts as an empty queue at the beginning of the algorithm, and stores the root nodes of the graph and subgraphs created during the iterative procedure.

In Figure~\ref{f:topo}, the weakest edge in $G$ mentioned in the middle gray-shaded box corresponds to the edge associating with an estimated regression coefficient that indicates the weakest association between two nodes connected by this edge among all associations between connected pairs of nodes. We use $p$-values of the estimated regression coefficients to identify the weakest edge to be removed until the sorting algorithm resumes due to newly emerging root nodes. By the time the queue $\mathscr{T}$ collects all $p$ nodes, we obtain a final regression coefficient matrix estimate by placing zeros in the entries corresponding to the removed weak edges. 

A complete algorithm for finding a minimizer of the penalized score-based objective function $R(\bB)$ in (\ref{eq:bigBhat}) that corresponds to a DAG is related next, which uses the PCD algorithm in conjunction with Kahn's sorting algorithm.  
\begin{description}
\item[Step 1:] Obtain an initial estimate of $\bB$ by solving $p$ unpenalized corrected score estimating equations one at a time. Denote by $\hat \bB^{(0)}$ the resultant initial estimate of $\bB$. Set the iteration index $t=0$.
\item[Step 2:] For $i\ne j=1, \ldots, p$, define $\tilde\beta_{ij}=\hat \bB^{(t)}[i,j]$ and $\tilde\beta_{ji}=\hat \bB^{(t)}[j, i]$. For each pair of nodes $i$ and $j$, update $(\tilde\beta_{ij}, \tilde\beta_{ji})$ to $(\tilde\beta_{ij}^*, \tilde\beta_{ji}^*)$ by minimizing the penalized score-based objective function following the algorithm elaborated in Appendix B of the Supplementary Materials. Set $\hat \bB^{(t+1)}=[\tilde\beta^*_{ij}]_{i,j=1, \ldots, p}$. Denote by $\tilde G$ the graph structure indicated by $\hat \bB^{(t+1)}$, which may not be a DAG. 
\item[Step 3:] For $j=1, \ldots, p$, compute unpenalized corrected score estimates for regression coefficients associated with the parents of $X_j$ suggested by $\tilde G$. Obtain estimated standard errors associated with these unpenalized regression coefficients estimates via sandwich variance estimation for M-estimators. Produce $p$-values based on the corrected score estimate for $\beta_{ij}$ and its estimated standard error for testing $H_0: \, \beta_{ij}=0$ versus $H_1: \, \beta_{ij}\ne 0$, if $X_i$ is a parent of $X_j$ in $\tilde G$. 
\item[Step 4:] Implement Kahn's sorting algorithm to eliminate cycles in $\tilde G$ by setting some (initially nonzero in Step 3) coefficients in $\hat \bB^{(t+1)}$ to be zero that have the largest $p$-values, unless $\tilde G$ from Step 3 is a DAG. 
\item[Step 5:] If $|\hat \bB^{(t+1)}-\hat \bB^{(t)}|_{\infty}> 10^{-4}$, set $t=t+1$, and return to Step 2. Otherwise, output $\hat \bB^{(t+1)}$ as a minimizer of $R(\bB)$ that corresponds to a DAG. Here, for a matrix $\bA$, $|\bA|_\infty$ denotes the largest entry of $\bA$ in absolute value.  
\end{description} 
One can follow a similar algorithm described above to find the miminizer of the naive penalized log-likelihood function $R_{\textrm{nv}}(\bB)$ in (\ref{eq:naiveobj}) that relates to a DAG. This is elaborated in Appendix C of the Supplementary Materials, where formulas for updating each pair of regression coefficients in Step 2 are provided. The algorithm implemented in \cite{Fu13} to minimize their penalized log-likelihood function using error-free data does not include Steps 3 and 4 above and thus does not guarantee to return a DAG in the end. 

When implementing the PCD algorithm, one essentially considers one pair of regression models at a time in each iteration, which are the $j$th and the $i$th regression models, that is, the regression model with $X_j$ as the response and the one with $X_i$ as the response, respectively. For each pair of models, one focuses on inferring one regression coefficient in each model in that iteration. In particular, one infers if $X_i$ should be included as an influential covariate in the $j$th regression model or if $X_j$ should be an influential covariate in the $i$th regression model, given all other covariates chosen from the previous iteration for that model. Alternatively, instead of updating $\hat\bB^{(t)}$ one pair of entries at a time, one may update one column of $\hat\bB^{(t)}$ at a time by selecting important covariates for the $j$th regression model, for $j=1, \ldots, p$. This leads to another approach for estimating $\bB$ that follows a similar algorithm but with the following step replacing Step 2 above: 
\begin{description}
\item[Step $2^*$:] For $j=1, \ldots, p$, use $\hat \bB_j^{(t)}$ as the starting value to solve the following penalized score estimating equation, 
\begin{equation}
n^{-1}_{-j}\sum_{\ell \in O_j} \bPsi_{j\ell}(\bB_j)-\tilde P_\lambda(\bB_j) =\bzero,
\label{eq:penest}
\end{equation}
where $\tilde P_\lambda(\bB_j)$ is a $(p-1)\times 1$ vector with entries given by, for $k\ne j$, 
$$\frac{\partial}{\partial \beta_{kj}} P_\lambda(|\beta_{kj}|)=\lambda \left\{I(|\beta_{kj}|\le \lambda)+\frac{(a\lambda-|\beta_{kj}|)_+}{(a-1)\lambda} \right\}\textrm{sign}(\beta_{kj}).$$
Let the resultant $p$ sets of solutions as the $p$ columns in the updated estimated $\bB$, $\hat \bB^{(t+1)}$. Denote by $\tilde G$ the graph structure indicated by $\hat \bB^{(t+1)}$.
\end{description}
We use Newton-Raphson algorithm to solve (\ref{eq:penest}), where the derivative of $\tilde P_\lambda(\bB_j)$ is approximated by a $(p-1)\times (p-1)$ diagonal matrix whose diagonal entries are given by $I(\beta_{kj}\ne 0)|(\partial/\partial \beta_{kj}) P_\lambda(|\beta_{kj}|)|/|\beta_{kj}|$,  for $k\ne j$. This is also the local quadratic approximation used in \cite{Fan01}. We refer to this algorithm as the node-wise parent selection (NPS) algorithm to distinguish from the previous algorithm that involves PCD. 

For each node, the NPS algorithm in Step $2^*$ is precisely the algorithm proposed by \cite{huang2013variable} for variable selection in one linear regression model with error-prone covariates. Consistency of this method for one regression model is established by the authors. Hence, without considering the correlation between $p$ regression models that the Bayesian network decomposes into, we expect that this alternative algorithm can yield a sensible estimate for $\bB$ that ignores the acyclicity constraint, and the cycle detection and elimination in Step 3 allows one to impose this constraint on the output of the NPS algorithm. Putting the penalty term aside, solving the $p$ sets of penalized score estimation equations in (\ref{eq:penest}) is intrinsically related to minimizing the penalized score-based objective function in (\ref{eq:bigBhat}) since they both originate from the corrected score. 

\section{Tuning Parameter Selection}
\label{s:chooselam}
We are now in the position to discuss choices of the tuning parameter $\lambda$ in the penalized  score-based objective function in  (\ref{eq:bigBhat}) and the penalized score estimating equation in (\ref{eq:penest}). In principle, it is desirable to use a consistent information criterion to choose $\lambda$. Assume that the class of candidate models includes a true model which the observed data come from, then a consistent information criterion refers to a criterion approaching (in probability) to its optimal value as the sample size tends to infinity when evaluated at the true model. In the context of variable selection in a regression model, within the class of all candidate models, a correct model includes all truly influential predictors in the true model and may also include non-influential predictors; the rest are incorrect models, which are referred to as underfitted models. In other words, the true model is the most parsimonious correct model, and a correct model that is not the true model is an overfitted model. Hence, with probability tending to one as the sample size increases, a consistent information criterion evaluated at the true model reaches its optimal value compared to when it is evaluated at an overfitted or underfitted model.   

To infer a Bayesian network consisting of error-prone nodes, we propose the {\underline s}core-based {\underline i}nformation {\underline c}riterion evaluated at a graph $G$ given by 
\begin{equation}
\textrm{SIC}(G)=\sum_{j=1}^p\left(\hat V_j+e_j\frac{\log n_{-j}}{n_{-j}}\right), \label{eq:SICGm}
\end{equation}
where $e_j$ is the number of parents of $X_j$ according to $G$, and $\hat V_j$ is equal to $V_j$ evaluated at the unpenalized corrected score estimate of $\bB_j$ given the structure of $G$. In the context of linear regression with error-prone covariates, \cite{huang2013variable} developed two score-based information criteria very much in the same spirit as the summand in (\ref{eq:SICGm}) to facilitate variable selection in one regression model. The proposed information criterion in (\ref{eq:SICGm}) is essentially the sum of $p$ score-based information criteria associated with $p$ regression models as the decomposition of a Bayesian network. To establish its consistency as a model criterion, it is instructive to relate arguments for model selection in the context of one regression model to arguments for graph selection, where a graph can be decomposed into $p$ regression models. 

Denote by $\bE_G$ the set of directed edges in $G$, and by $|\bE_G|$ the size of this set.
Suppose there exists a true graph $G_0$ in the class of graphs under consideration, which dictates the data generating process. Parallel with notions in variable selection in the regression setting, let $G_-$ and $G_+$ denote generically an underfitted graph and an overfitted graph, respectively, where $G_-$ satisfies $\bE_{G_0}\not\subset \bE_{G_-}$, and $G_+$ satisfies $\bE_{G_0}\subset \bE_{G_+}$. Then $G_0$ and $G_+$ are correct graphs, with the former more parsimonious than the latter, that is, $|\bE_{G_0}|<|\bE_{G_+}|$. In contrast, $G_-$ is an incorrect graph, and one does not necessarily have $|\bE_{G_-}|<|\bE_{G_0}|$. To establish the consistency of $\textrm{SIC}(G)$, it suffices to show that
\begin{eqnarray*}
\textrm{SIC}(G_-)-\textrm{SIC}(G_0) & > 0 & \textrm{ with probability approaching one, as $n\to \infty$ and}, \label{eq:underfitm}  \\
\textrm{SIC}(G_+)-\textrm{SIC}(G_0) & \to 0^+ & \textrm{ in probability as $n\to \infty$,} \label{eq:overfit} 
\end{eqnarray*}
where $n=\min_{1\le j \le p}n_{-j}$. These assertions are proved in Appendix D of the Supplementary Materials, where $p$ is allowed to diverge as $n\to \infty$.

\section{Empirical Evidence}
\label{s:simulation}
\subsection{Competing Methods}
In this section, we implement the proposed score-based methods and the naive likelihood-based method using simulated network data to assess their finite sample performance. 
For the score-based methods, we use the SIC tuning parameter selector to choose $\lambda$. 
For the naive method, we adopt the tuning parameter selection method employed in \cite{Fu13} based on the relative change in the prediction error. 

Denote by $e_\lambda$ the number of edges of an estimated graph when the tuning parameter is set at $\lambda$, and by $\hat\bB_{\textrm{nv}}^{(\lambda)}$ the corresponding naive estimate of $\bB$. Define the prediction error by $\textrm{PE}_{\lambda}=\sum_{j=1}^p \sum_{\ell \in O_j}(\bW[\ell, j]-\hat\bW^{(\lambda)}[\ell, j])^2,$ where $\hat\bW^{(\lambda)}[\ell, j]=\bW[\ell, -j]\hat\bB_{\textrm{nv},j}^{(\lambda)}$.  Suppose one considers $m$ candidate values for $\lambda$, $\lambda_1>\lambda_2>\ldots>\lambda_m$. For each $k=2, \ldots, m$, one computes the {\underline r}elative {\underline c}hange in {\underline p}rediction error defined by $\textrm{RCP}_{k-1, k}=(\textrm{PE}_{\lambda_{k-1}}-\textrm{PE}_{\lambda_k})/(e_{\lambda_k}-e_{\lambda_{k-1}})$, if $e_{\lambda_k}-e_{\lambda_{k-1}}>0$, and $\textrm{RCP}_{k-1, k}=0$ otherwise. Then one chooses $\lambda_K$ as the tuning parameter value, where $K=\max\{k: \textrm{RCP}_{k-1,k}\ge \alpha \max(\textrm{RCP}_{1,2}, \ldots, \textrm{RCP}_{m-1, m}), \, k=2, \ldots, m\}$, in which $\alpha$ is a threshold parameter set at 0.1 in our simulation study. The quantity defined as $\textrm{RCP}_{k-1, k}$ essentially quantifies how much one gains in prediction accuracy at the price of increasing graph complexity (as $e_\lambda$ increases) when one drops $\lambda$ from $\lambda_{k-1}$ to $\lambda_k$. The use of the threshold $\alpha$ is to further guard again overly dense graphs. The constructions of RCP and $K$ together aim to balance graph complexity and prediction accuracy. 

In summary, there are three methods implemented in the simulation study: the naive likelihood-based method using the PCD algorithm with $\lambda$ chosen by RCP, the score-based method using the PCD algorithm with $\lambda$ chosen by SIC, and the score-based method using the NPS algorithm with $\lambda$ chosen by SIC. %For notational simplicity, we code these three methods as NV-PCD, CS-PCD, and CS-NPS, respectively.

\subsection{Simulation Settings}
The simulation experiment involves two factors: the number of nodes $p$ and the variance-covariance matrix of the measurement error $\bSigma_u$. There are two levels for $p$, 10 and 20. Given $p$, the total number of edges of a true graph is set to be $3p$, and each node has at most four parents. Once such a graph is created randomly, we set the entries in $\bB$ associated with the first half of edges at 0.5, and entries associated with the second half of edges at 1. Then we generate $n_j=5$ interventional data points from $N(0, 1)$, for $j=1, \ldots, p$. When generating normal measurement errors, we first set $\bSigma_u=\sigma_u^2 \bI_p$, where $\sigma_u^2$ varies across 5 levels to produce reliability ratio associated with each $X_j$, defined by $\tau=\textrm{Var}(X_j)/\{\textrm{Var}(X_j)+\sigma_u^2\}$, ranging from 0.8 to 1 at increments of 0.05, for $j=1, \ldots, p$. In a different setting we let $\bSigma_u=\sigma_u^2 \bV_p$, where $\sigma^2_u$ takes the five aforementioned levels, and $\bV_p$ is a $p\times p$ matrix with entries given by $\bV_p[j, j']=0.5^{|j-j'|}$, for $j, j'=1, \ldots, p$. For each simulation setting, we randomly generate ten graphs, from each of which an $N\times p$ data matrix $\bW$ is generated according to (\ref{eq:regwx}) and (\ref{eq:wxadd}) with $\{\bepsilon[\ell, j], \ell=1, \ldots, N\}_{j=1}^p$ being independent realizations from $N(0, 1)$. 

Given a true graph $G$, the following five metrics are used to assess the quality of an estimated graph $\hat G$: the true positive rate, TPR $=|\bE_{\hat G}\cap\bE_{G}|/(3p)$; the false discovery rate, FDR $=(\textrm{R}+|\bE_{\hat G}\cap\bE_{G}^c|)/|\bE_{\hat G}|$, where R denotes the number of edges in $G$ that show up in $\hat G$ in the reversed direction; the specificity $=|\bE^c_{\hat G}\cap\bE_{G}^c|/\{p(p-4) \}$, where $p(p-4)=p^2-p-3p$ is the number of zero non-diagonal entries in $\bB$; the rate of correct identification of existence (with the right direction) and non-existence of edges defined as $(|\bE_{\hat G}\cap\bE_{G}|+|\bE^c_{\hat G}\cap\bE_{G}^c|)/\{p(p-1)/2\}$; and lastly, the Frobenius norm of $\bB-\hat \bB$ divided by the number of off-diagonal entries of $\bB$, that is, $\textrm{trace}\{(\bB-\hat \bB )(\bB-\hat \bB)^\T\}/\{p(p-1)\}$. The first four metrics are of interest when one is concerned about inference on the graph structure, and the last metric is of interest when one wishes to understand the strength of associations between nodes, and to use the estimated graph for prediction.

\subsection{Simulation Results}
\begin{landscape}
\begin{figure}
\centering
\setlength{\linewidth}{1.2\textwidth}
	\includegraphics[width=\linewidth]{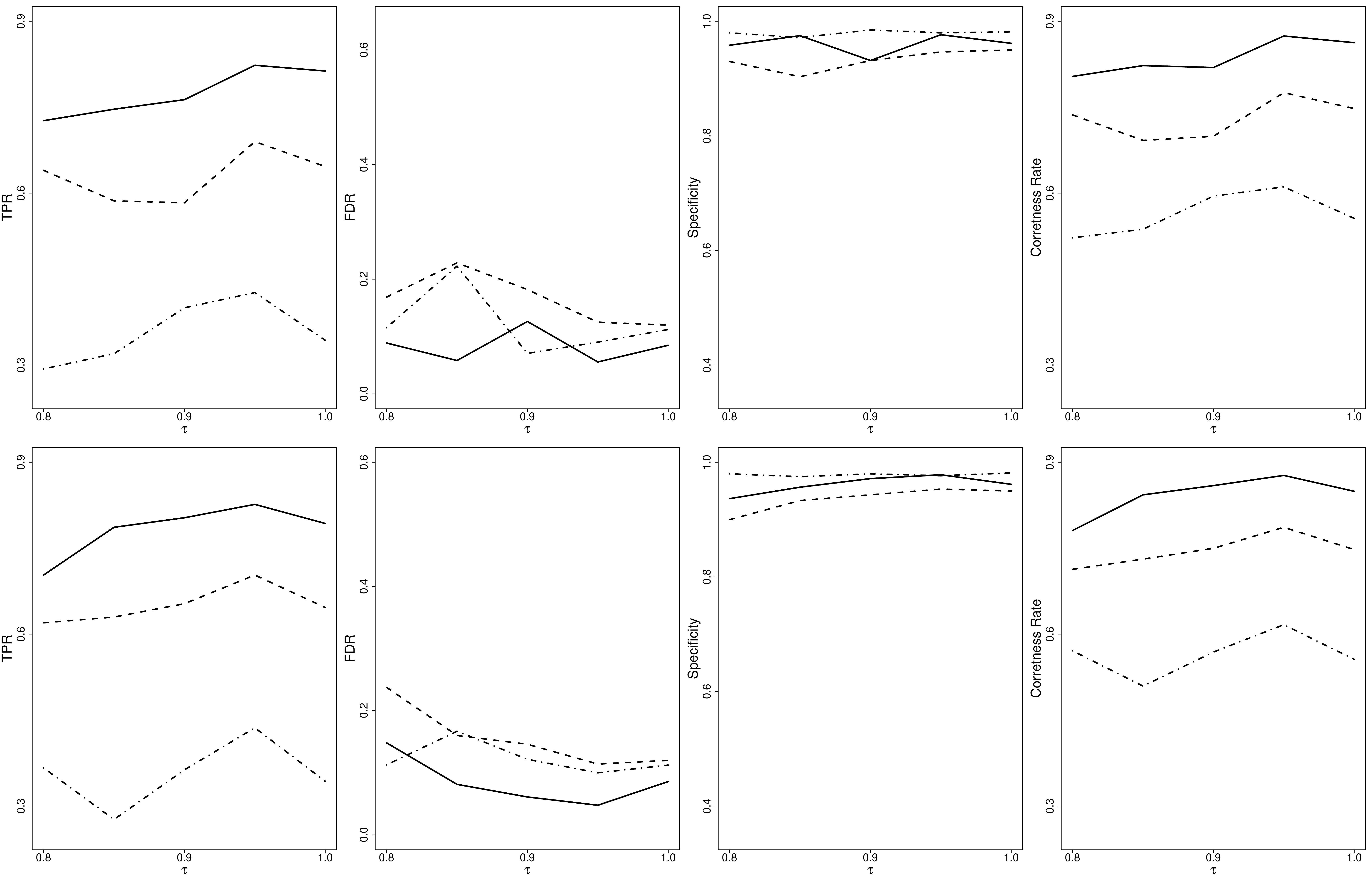}
\caption{\label{f:p10} Monte Carlo averages of TPR, FDR, specificity, and correctness rate versus the reliability ratio $\tau$ across ten graphs with $p=10$ nodes associated with three methods, the method by \cite{Fu13} (dash-dotted lines), corrected score method using PCD algorithm (dashed lines), and corrected score method using NPS algorithm (solid lines), when $\bSigma_u$ is a diagonal matrix (top panels) and when it is not a diagonal matrix (bottom panels).}
\end{figure}
\end{landscape}

\begin{landscape}
\begin{figure}
\centering
\setlength{\linewidth}{1.2\textwidth}
	\includegraphics[width=\linewidth]{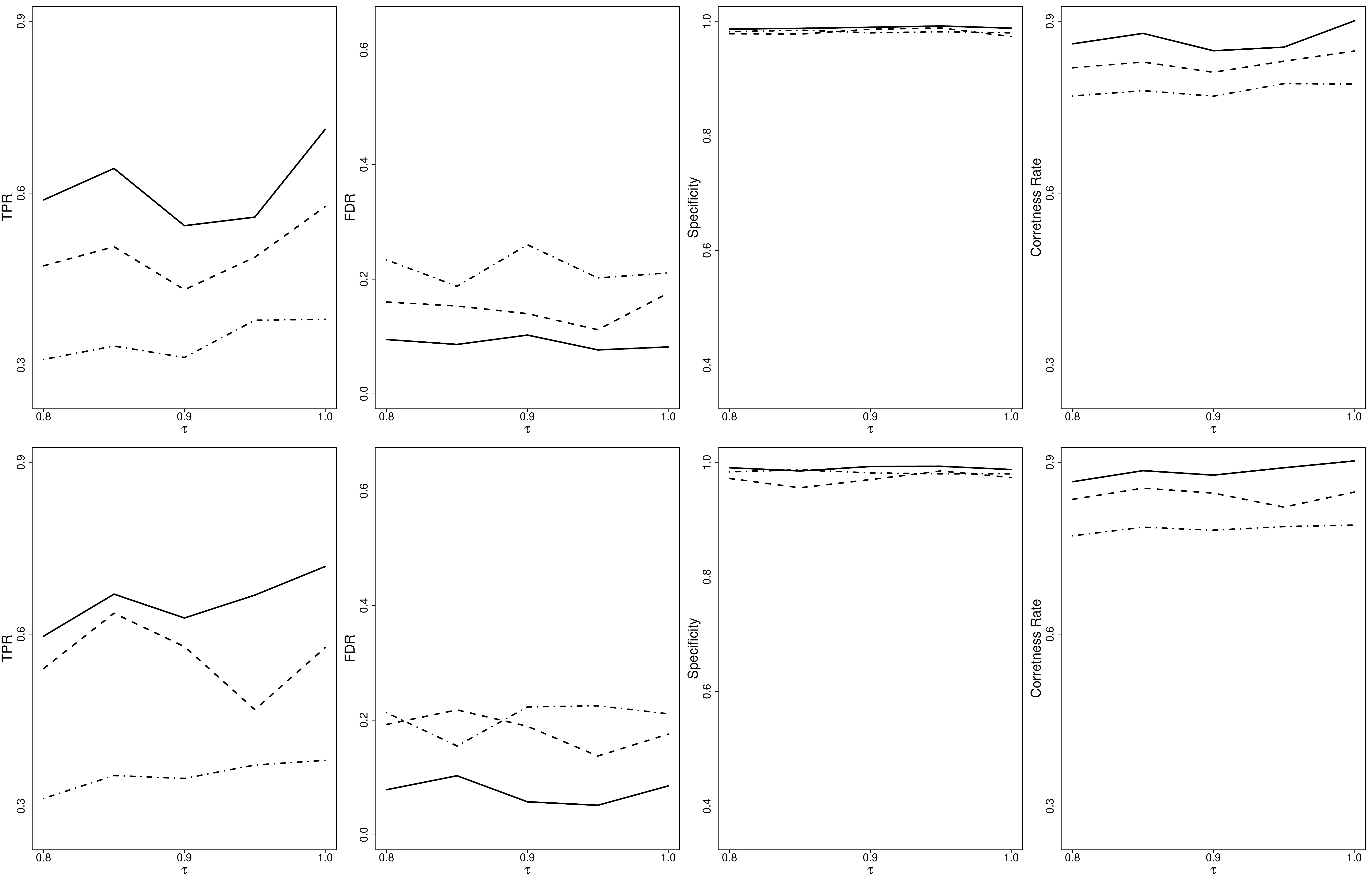}
\caption{\label{f:p20} Monte Carlo averages of TPR, FDR, specificity, and correctness rate versus the reliability ratio $\tau$ across ten graphs with $p=20$ nodes associated with three methods, the method by \cite{Fu13} (dash-dotted lines), corrected score method using PCD algorithm (dashed lines), and corrected score method using NPS algorithm (solid lines), when $\bSigma_u$ is a diagonal matrix (top panels) and when it is not a diagonal matrix (bottom panels).}
\end{figure}
\end{landscape}

Figure~\ref{f:p10} depicts the Monte Carlo (MC) averages (across ten graphs) of TPR, FDR, specificities, and rates of correct identification of existence/non-existence of directed edges associated with three considered methods when $p=10$ under two specifications of $\bSigma_u$. Figure~\ref{f:p20} shows the same collection of results when $p=20$. Across these four metrics, the advantages of the score-based methods pairing with the SIC tuning parameter selector are evident over a wide range of reliability ratio $\tau$, whether the PCD algorithm is used for implementing the corrected score method, or the NPS algorithm is used. The naive likelihood-based method suffers from low TPR, although it is comparable with the score-based methods in terms of specificity. This phenomenon  can be explained by the well-known attenuation effect of measurement error on slope parameters estimates in a linear regression model with classical measurement error in covariates \citep[][Section 1.1]{fuller2009measurement}. More specifically, in the context of linear regression with covariates measurement error, naive estimators of covariate effects tend to  attenuate towards zero, which explains the low TPR. Such attenuation effect does not compromise naive estimation of a null covariate effect, which explains the robustness of specificity to measurement error. As a combination of TPR and specificity, the correction rate observed for the naive method is also less affected by measurement error than TPR alone. This robustness is more evident in a sparser graph, such as a graph consisting of $p=20$ nodes with $3p$ edges when comparing with a graph consisting of $p=10$ nodes with $3p$ edges. Here, a measure of sparsity of a graph $G$ can be defined as $|\bE_G|/\{p(p-1)/2\}$, where $p(p-1)/2$ is the largest number of edges possible for a DAG with $p$ nodes. Finally, even in the absence of measurement error (i.e., with $\tau=1$ in Figures~\ref{f:p10} and \ref{f:p20}), the two score-based methods still outperform the likelihood-based method when TPR and correction rate are considered. This implies that the construction of the (unpenalized) objective function plays an important role in network inference. 

\begin{figure}
\centering\setlength{\linewidth}{0.8\textwidth}
	\includegraphics[width=\linewidth]{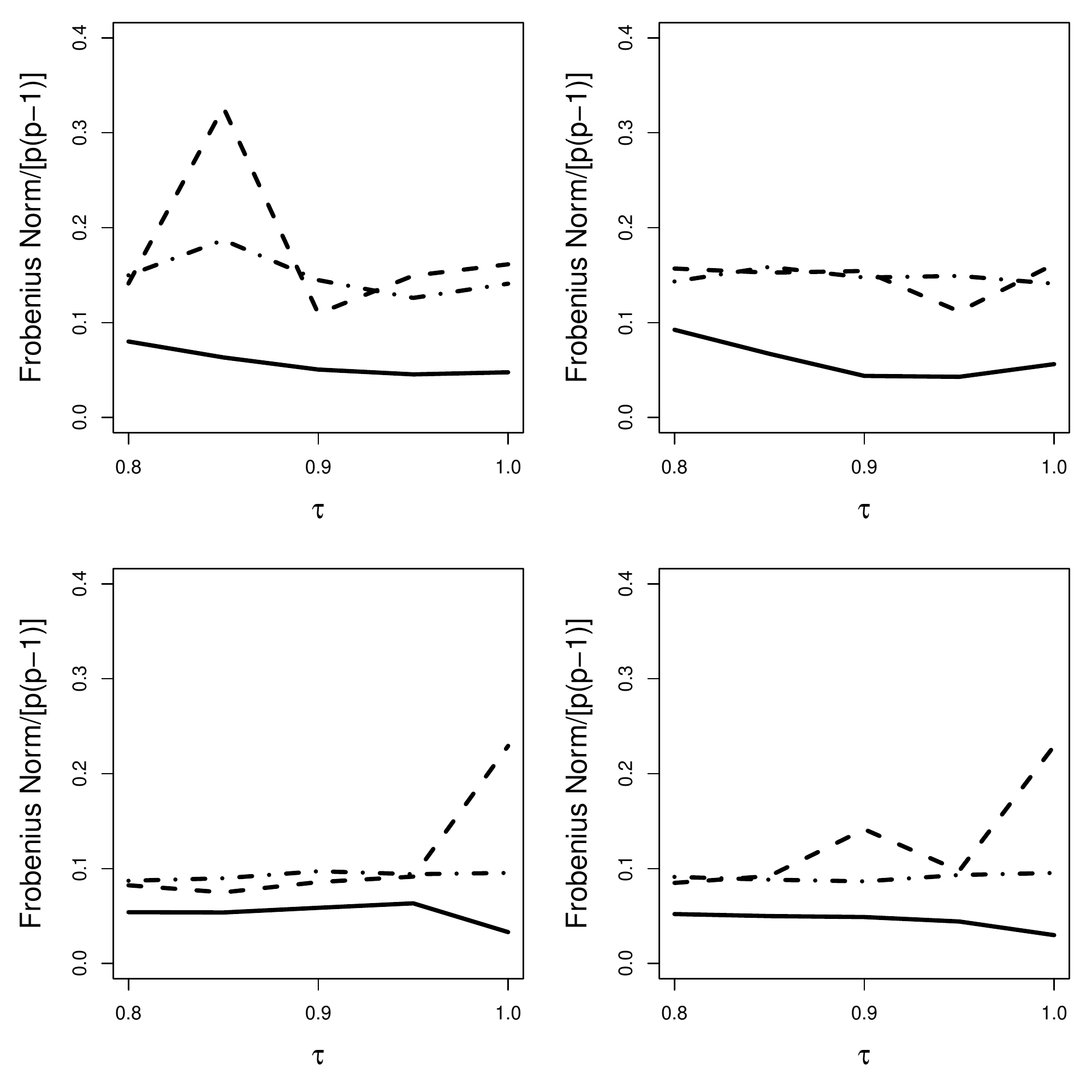}
\caption{\label{f:fnorm} Monte Carlo medians of the Frobenius norm of $\bB-\hat \bB$ divided by $p(p-1)$ versus the reliability ratio $\tau$ across ten graphs with $p=10$ nodes (top panels) and $p=20$ nodes (bottom panels) associated with three methods, the method by \cite{Fu13} (dash-dotted lines), corrected score method using PCD algorithm (dashed lines), and corrected score method using NPS algorithm (solid lines), when $\bSigma_u$ is a diagonal matrix (left panels) and when it is not a diagonal matrix (right panels).}
\end{figure}

Figure~\ref{f:fnorm} shows MC medians of the Frobenius norm of $\bB-\hat \bB$ divided by $p(p-1)$. This figure suggests that the PCD algorithm can lead to some numerical instability for the corrected score method, and the NPS algorithm produces more stable regression coefficients estimates from the corrected score method that are also less biased than the naive estimates. In fact, between the two score-based methods, the one using the NPS algorithm yields better inference outcomes in all aspects depicted in Figures~\ref{f:p10}--\ref{f:fnorm} than those resulting from the PCD algorithm. This suggests that there may exist some interaction effect of regression coefficients estimation and cycle elimination procedure on the finite sample performance of a method. 

\subsection{Application to Flow Cytometry Data}
\label{s:realdata}
Now we return to the application of inferring cellular signaling networks using flow cytometry data. In particular, the flow cytometry data we entertain in this section consist of $p=11$ phosphomolecular measurements from each of $N=7466$ human immune system cells collected in an experiment described in \cite{Sachs05}. In this experiment, a series of stimulatory cues and inhibitory interventions were imposed, producing the observed data matrix as a mixture of observational data and interventional data for the eleven phosphorylated proteins and phospholipids \citep[see Table 1 in][]{Sachs05}. \cite{Shojaie10} applied a penalized likelihood estimation method with LASSO and ALASSO penalty to infer the signaling network while assuming ordering of the eleven nodes known. Without assuming ordering known,  \cite{Fu13} applied their likelihood-based penalized estimation method on this data set to infer a directed signaling network using the PCD algorithm, also treating the data as measures of the true nodes. \cite{luo2011bayesian} viewed the observed data as error-contaminated surrogates of the true protein activity levels, and assumed a normal additive measurement error, with an inverse gamma prior distribution for the measurement error variance (common for all nodes). Neither of the aforementioned methods  guarantees that the inferred graph is acyclic. 

Because this data do not contain replicate measures of the same underlying protein activity level, error variance is not identifiable, even with the normality assumption imposed. A widely adopted practice in the measurement error literature in this case is to carry out sensitivity analysis, where one assumes different values for the error variance to observe how inference results from a considered method vary. This exercise can be helpful for addressing the robustness of a method to the misspecification of measurement error variance. For the purpose of comparing our proposed score-based methods that account for measurement error with the naive likelihood-based method that ignores measurement error in nodes, we follow the viewpoint in \cite{Fu13} and treat the observed phosphomolecular measurements as the intended (error-free) measures of the true nodes, whose ordering is unknown. This allows us to have benchmark inferences, based on which we are able to compare results from the two proposed methods with those from the naive method, all three applied to error-contaminated data.

We apply our score-based penalized estimation methods based on error-prone data $\bW$ generated from contaminating $\bX$ according to (\ref{eq:wxu}) with an estimated reliability ratio of 0.8, where the variance of each node is estimated by its interventional data. The computer code for this data analysis along with the data are available in the supplementary materials. Panel (a) in Figure~\ref{f:realdata} shows a network with directed edges reflecting causal relationships between these nodes that are currently well accepted in the literature. Networks shown in panels (b)--(f) in Figure~\ref{f:realdata} include the one from \cite{Shojaie10} using the ALASSO penalty and assuming data free of measurement error with ordering known, the network from \cite{Fu13} applied to $\bX$, the naively inferred network based on $\bW$, and two networks obtained from the corrected score methods, implemented via the PCD algorithm and via the NPS algorithm, respectively. When comparing each of the latter five networks with the consensus graph, the network from \cite{Shojaie10} includes 14 edges in the consensus network among a total of 27 edges in their inferred graph; and there are 8 edges in the consensus network included in the network from \cite{Fu13}, which also has a total of 27 edges. When error-prone data are used (see panels (d)--(f)), the naive method produces a very sparse graph, with merely 8 edges, among which 4 are in the consensus graph; the corrected score method implemented via the PCD algorithm leads to a much denser graph, with 37 edges, 9 of which are in the consensus graph; the corrected score method using the NPS algorithm results in a graph with 28 edges, 10 of which are in the consensus graph.

Using the consensus graph as a gold standard, the above comparisons between the six networks suggests that, when error-prone data are used for inferring a Bayesian network, the naive likelihood-based method can lead to low discovery rate, and the corrected score methods can identify more truly existing causal relationships between nodes, although using the PCD algorithm can result in a higher false discovery rate than when the NPS algorithm is used. 

\section{Discussion}
\label{s:discussion}
We proposed score-based methods to infer a Bayesian network using error-prone data from interventional experiments. When only observational data are available, the proposed method can be used to infer graphs within a Markov equivalence class \citep{andersson1997characterization} since a graph is not identifiable using observational data only but a Markov equivalence class is. A consistent model criterion is also constructed based on the same score function for tuning parameter selection. Besides establishing the consistency in the resulting regression coefficients estimator, we also provide convincing empirical evidence to show that the proposed score-based methods can substantially outperform a naive likelihood-based method that ignores measurement error. And, even in the absence of measurement error in nodes, using a quadratic inference function constructed based on an unbiased score is more preferable than using a likelihood function to formulate a penalized objective function for network estimation. We exploit Kahn's topological sorting algorithm along with the PCD algorithm or the NPS algorithm to estimate the regression coefficients matrix, which are computationally less burdensome than  many search-and-score methods that attempt to select a graph from a DAG family of size that grows super-exponentially fast as $p$ grows \citep{Robinson1973}. One computational hurdle remains for the proposed method when $p$ is large is the inversion of a $(p-1)\times (p-1)$ matrix in (\ref{eq:nonpenobj}). A model criterion that does not involve the inversion of a large matrix is more desirable in that case. 

It is assumed that both model error in (\ref{eq:DAGx}) and measurement error in (\ref{eq:wxu}) are Gaussian. When the normality assumption is violated, the corrected score in (\ref{eq:sumscorej}) may not be an unbiased score. Constructing score functions that are robust to the normality assumption and also account for measurement error is a follow-up research direction. This is also the direction one can follow to relax the linearity assumption of the regression model in (\ref{eq:DAGx}).

\begin{landscape}
\begin{figure}
\centering
\setlength{\linewidth}{1.2\textwidth}
\includegraphics[width=\linewidth]{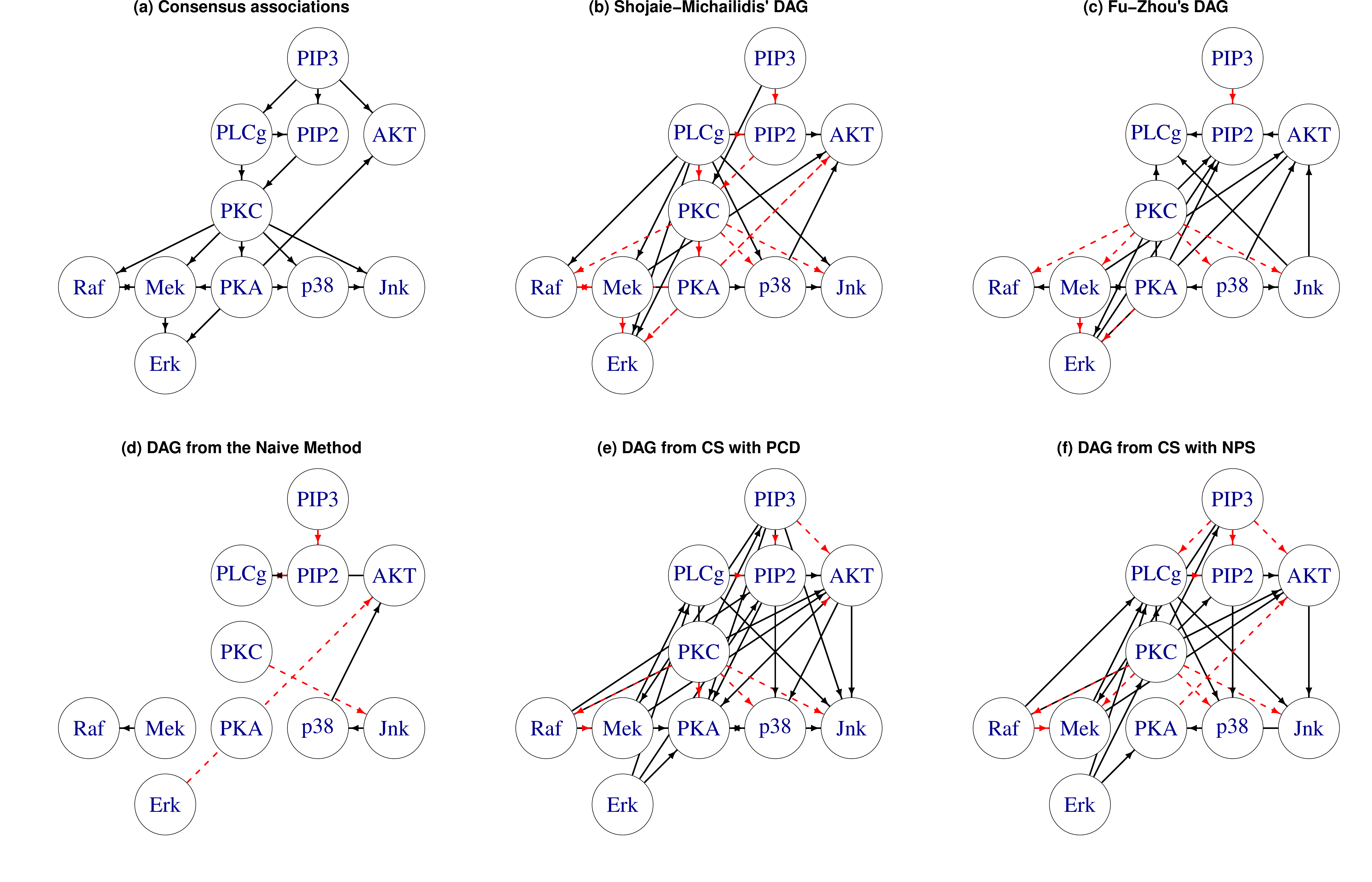}
\caption{\label{f:realdata} Six signaling networks associated with the flow cytometry data set: (a) the consensus graph, (b) the estimated graph from \cite{Shojaie10} assuming ordering known, (c) the estimated graph based on $\bX$ from \cite{Fu13}, (d) the estimated graph based on $\bW$ using the naive method, (e) the estimated graph based on $\bW$ using the corrected score method and PCD algorithm, (f) the estimated graph based on $\bW$ using the corrected score method and NPS algorithm. In graphs (b)--(f), the inferred edges in agreement with (a) are highlighted as red dashed edges.}
\end{figure}
\end{landscape}

\section*{Appendix A: Proof of Theorem 3.1}
\setcounter{section}{0}
\renewcommand{\theequation}{A.\arabic{equation}}
Denote by $\bB^*$ the true value of $\bB$. Define $\bb=(\bB_1^\T, \ldots, \bB_p^\T)^\T$, where $\bB_j=\bB[-j, j]$, for $j=1, \ldots, p$; $\hat\bb$ and $\bb^*$ are similarly defined. In this appendix, we prove the following theorem. 

\noindent
{\bf Theorem 3.1} Under assumptions (A1)--(A5), as $n=\min_{1\le j \le p} n_{-j} \to \infty$, if $\sqrt{n}\lambda_n=o_p(1)$, then there exists a local minimizer of $R(\bB)$ defined in equation (8) in the main article, denoted by $\hat \bB$, such that $\|\hat\bb-\bb^*\|=O_p(n^{-1/2})$.

For $j=1, \ldots, p$, we impose the following assumptions, 
\begin{description}
\item[(A1)] the truth, $\bB^*_j$, is a solution to $\lim_{n_{-j}\to \infty} n_{-j}^{-1}\sum_{\ell \in O_j} E\{\bPsi_{j\ell}(\bB_j)\}=\bzero$;
\item[(A2)] $\lim_{n_{-j}\to \infty} n_{-j}^{-1}\sum_{\ell \in O_j} E\{\bPsi_{j\ell}(\bB^*_j)\bPsi^t_{j\ell}(\bB^*_j)\}$ exists and is positive definite;
\item[(A3)] $\lim_{n_{-j}\to \infty} n_{-j}^{-1}\sum_{\ell \in O_j} E\{-(\partial/\partial \bB_j^t)\bPsi_{j\ell}(\bB_j)\vert_{\bB_j=\bB_j^*}\}$ exists and is positive definite; 
\item[(A4)] $\lim_{n_{-j}^{-1}\to \infty} n_{-j}\sum_{\ell \in O_j} (\partial/\partial \bB_j^\T)\{\bH^{-1}_j(\bB_j)\bpsi_{j\ell}(\bB_j)\}\big\vert_{\bB_j=\bB_j^*}$ exists; 
\item[(A5)] $E\{V_j^{(2)}(\bB_j^*)\}$ is positive definite with eigenvalues uniformly bounded by a positive constant.  
\end{description}

Recall that the penalized objective function is
$$R(\bB)=\sum_{j=1}^p \left\{V_j(\bB_j)+\sum_{i=1}^p P_{\lambda_n}(|\beta_{ij}|) \right\}.$$
To show $\|\hat\bb-\bb^*\|=O_p(n^{-1/2})$, it suffices to show that, for any $\epsilon>0$, there exists a large enough positive constant $C$ such that 
\begin{equation}
P\left\{\inf_{\|\textrm{vec}(\bu)\|=C} R(\bB^*+n^{-1/2}\bu)>R(\bB^*)\right\}\ge 1-\epsilon, \label{eq:suffcond}
\end{equation}
where $\bu$ is a non-random $p\times p$ matrix with zeros on the diagonal, and $\textrm{vec}(\bu)=(\bu_1^\T, \ldots, \bu_p^\T)^\T$. 

Denote by $\Pi_j=\{i\in \{1, \ldots, p\}: \beta_{ij}^*\ne 0\}$, that is, $\Pi_j$ is the index set corresponding to the parents of $X_j$, for $j=1, \ldots, p$. By the definition of $R(\bB)$, we have
\begin{eqnarray}
& & R(\bB^*+n^{-1/2}\bu)-R(\bB^*) \nonumber \\
& = & \sum_{j=1}^p \left\{V_j(\bB_j^*+n^{-1/2} \bu_j)-V_j(\bB_j^*)\right\}+\sum_{j=1}^p\sum_{i=1}^p\left\{P_{\lambda_n}(|\beta_{ij}^*+n^{-1/2}u_{ij}|)-P_{\lambda_n}(|\beta_{ij}^*|)\right\} \nonumber \\
& = & \sum_{j=1}^p \left[n^{-1/2}\left\{V^{(1)}_j(\bB_j^*)\right\}^\T \bu_j+0.5 n^{-1} \bu_j^\T V^{(2)}_j(\bB_j^*)\bu_j\{1+o_p(1)\}\right]+ \nonumber \\
&  & \sum_{j=1}^p\left[\sum_{i \in \Pi_j} \left\{P_{\lambda_n}(|\beta_{ij}^*+n^{-1/2}u_{ij}|)-P_{\lambda_n}(|\beta_{ij}^*|)\right\} + \sum_{i \notin \Pi_j} \left\{P_{\lambda_n}(|\beta_{ij}^*+n^{-1/2}u_{ij}|)-P_{\lambda_n}(|\beta_{ij}^*|)\right\}\right], \nonumber
\end{eqnarray}
where we apply the second order Taylor expansion of $V_j(\bB_j^*+n^{-1/2} \bu_j)$ around $\bB_j^*$ in the first sum above; and, since $\beta^*_{ij}=0$ for $i\notin \Pi_j$, the second sum is equal to 
\begin{eqnarray*}
& & \sum_{j=1}^p\left[\sum_{i \in \Pi_j} \left\{P_{\lambda_n}(|\beta_{ij}^*+n^{-1/2}u_{ij}|)-P_{\lambda_n}(|\beta_{ij}^*|)\right\} + \sum_{i \notin \Pi_j} P_{\lambda_n}(|n^{-1/2}u_{ij}|)\right] \nonumber \\
& \ge & \sum_{j=1}^p \sum_{i \in \Pi_j} \left\{P_{\lambda_n}(|\beta_{ij}^*+n^{-1/2}u_{ij}|)-P_{\lambda_n}(|\beta_{ij}^*|)\right\} \nonumber \\
& \ge & \sum_{j=1}^p \sum_{i \in \Pi_j} P'_{\lambda_n}(|\beta_{ij}^*+n^{-1/2}u_{ij}|)(|\beta_{ij}^*+n^{-1/2}u_{ij}|-|\beta_{ij}^*|), \nonumber
\end{eqnarray*}
in which the last inequality is due to the concavity of $P_{\lambda_n}(t)$ on $[0, \infty)$. It follows that, for a large enough $n$, 
\begin{eqnarray}
& & R(\bB^*+n^{-1/2}\bu)-R(\bB^*) \nonumber \\
& \ge & \sum_{j=1}^p \Big[n^{-1/2}\left\{V^{(1)}_j(\bB_j^*)\right\}^\T \bu_j+0.5 n^{-1} \bu_j^\T V^{(2)}_j(\bB_j^*)\bu_j\{1+o_p(1)\}+ \nonumber \\
& & \sum_{i \in \Pi_j} P'_{\lambda_n}(|\beta_{ij}^*+n^{-1/2}u_{ij}|)n^{-1/2}u_{ij}\textrm{sgn}(\beta_{ij}^*)\Big], \label{eq:threeterms}
\end{eqnarray}
where, within the summand, since $P'_{\lambda_n}(t)=O_p(\lambda_n)$, the third term is of order $o_p(n^{-1})$; and by assumption (A5), the second term is bounded from below by a term of the same order as $0.5n^{-1}v_jC$, where $v_j$ is the smallest eigenvalue (which is positive) of the limit of $\bV^{(2)}_j(\bB_j^*)$ as $n\to \infty$. As for the first term of the summand in (\ref{eq:threeterms}), we have
\begin{eqnarray*}
& & n^{-1/2}\left\{V^{(1)}_j(\bB_j^*)\right\}^\T \bu_j \\
& = & n^{-1/2}n_{-j}^{-1} \sum_{\ell \in O_j} \bpsi_{j\ell}^\T(\bB^*_j) \left[\bH_j^{-1}(\bB_j^*)n_{-j}^{-1}\sum_{\ell\in O_j}(\partial/\partial \bB_j^\T)\bpsi_{j\ell}(\bB_j)\big\vert_{\bB_j=\bB_j^*}\right.+\\
& & \left.n_{-j}\sum_{\ell \in O_j} (\partial/\partial \bB_j^\T)\{\bH^{-1}_j(\bB_j)\bpsi_{j\ell}(\bB_j)\}\big\vert_{\bB_j=\bB_j^*}\right]\bu_j,
\end{eqnarray*}
of which, by assumptions (A2)--(A4), the terms inside the square brackets altogether converge in probability to a bounded squared matrix; and, under (A1)--(A3), $\| n_{-j}^{-1} \sum_{\ell \in O_j} \bpsi_{j\ell}(\bB^*_j) \| = O_p(n_{-j}^{-1/2})=O_p(n^{-1/2})$. Hence $n^{-1/2}\left\{V^{(1)}_j(\bB_j^*)\right\}^\T \bu_j=O_p(n^{-1})$.

Combining the discussions on the three terms in (\ref{eq:threeterms}), we have that, as $n\to \infty$, for a large enough $C$, the second term in (\ref{eq:threeterms}) dominates the first and the third terms, thus $R(\bB^*+n^{-1/2}\bu)-R(\bB^*)>0$ in probability. This proves (\ref{eq:suffcond}) and thus Theorem 3.1. 
  
\setcounter{equation}{0}	
\section*{Appendix B: The PCD algorithm for the score-based method}
\setcounter{section}{0}
\renewcommand{\theequation}{B.\arabic{equation}}

Entering Step 2 in the algorithm in Section 3.2 of the main article, one implements the PCD algorithm to update one pair of regression coefficients $(\beta_{ij}, \beta_{ji})$ at a time by minimizing the penalized score-based objective function with all other entries in $\bB$ fixed. More specifically, for $i\ne j=1, \ldots, p$, define $\tilde\beta_{ij}=\hat \bB^{(t)}[i,j]$ and $\tilde\beta_{ji}=\hat \bB^{(t)}[j, i]$, one uses the following algorithm to update $(\tilde\beta_{ij}, \tilde\beta_{ji})$ to $(\tilde\beta_{ij}^*, \tilde\beta_{ji}^*)$:
\begin{itemize}
	\item[]PCD-1: Find 
	\begin{equation}
	\tilde\beta_{ij}^*=\argmin_{\beta_{ij}}\left\{\tilde V_j+\sum_{k\ne i,j}P_{\lambda}(|\tilde\beta_{kj}|)+P_		\lambda(|\beta_{ij}|)\right\},
	\label{eq:bij*}
	\end{equation}
	where 
	\begin{equation}
	\tilde V_j=\left\{n_{-j}^{-1} \sum_{\ell\in O_j} \bPsi_{j\ell}(\tilde\bB^*_j) \right\}^t \{\bH_j(\tilde\bB^*_j)\}^{-1} \left\{n_{-j}^{-1} \sum_{\ell\in O_j} \bPsi_{j\ell}(\tilde\bB^*_j) \right\},\label{eq:Vj}
	\end{equation}
	in which $\tilde\bB^*_j$ is the same as $\hat\bB^{(t)}_j$ except that $\tilde \beta_{ij}$ in $\hat\bB^{(t)}_j$ is replaced by $\beta_{ij}$. Note that $\beta_{ij}$ appears in both $\tilde V_j$ and $P_\lambda(|\beta_{ij}|)$.
	\item[]PCD-2: Find 
	\begin{equation}
	\tilde\beta_{ji}^*=\argmin_{\beta_{ji}}\left\{\tilde V_i+\sum_{k\ne i,j}P_{\lambda}(|\tilde\beta_{ki}|)+P_\lambda(|\beta_{ji}|)\right\},
	\label{eq:bji*}
	\end{equation}
	where $\tilde V_i$ is similarly defined as $\tilde V_j$ in (\ref{eq:Vj}). Note that $\beta_{ji}$ appears in both $\tilde V_i$ and $P_\lambda(|\beta_{ji}|)$.
	\item[]PCD-3: Compute 
	$$
	S_1  = \left\{\tilde V_i|_{\beta_{ji}=0}+\sum_{k\ne i,j}P_{\lambda}(|\tilde\beta_{ki}|)\right\}+
	\left\{\tilde V_j|_{\beta_{ij}=\tilde\beta^*_{ij}}+\sum_{k\ne i,j}P_{\lambda}(|\tilde\beta_{kj}|)+P_\lambda(|\tilde\beta^*_{ij}|)\right\},$$ 
	$$
	S_2 = \left\{\tilde V_i|_{\beta_{ji}=\tilde\beta^*_{ji}}+\sum_{k\ne i,j}P_{\lambda}(|\tilde\beta_{ki}|)+P_\lambda(|\tilde\beta^*_{ji}|)\right\}+\left\{\tilde V_j|_{\beta_{ij}=0}+\sum_{k\ne i,j}P_{\lambda}(|\tilde\beta_{kj}|)\right\},$$
	where, in $S_1$, $\tilde V_j|_{\beta_{ij}=\tilde\beta_{ij}^*}$ is $\tilde V_j$ in (\ref{eq:Vj}) with $\beta_{ij}$ evaluated at $\tilde\beta_{ij}^*$ from PCD-1, and $\tilde V_i|_{\beta_{ji}=0}$ is $\tilde V_i$ with $\beta_{ji}$ evaluated at zero. In $S_2$, $\tilde V_i|_{\beta_{ji}=\tilde\beta^*_{ji}}$ and $\tilde V_j|_{\beta_{ij}=0}$ are similarly defined. 
	\item[]PCD-4: If $S_1\le S_2$, then update $(\tilde\beta_{ij}, \tilde\beta_{ji})$ to $(\tilde\beta_{ij}^*, \, 0)$; otherwise, update $(\tilde\beta_{ij}, \tilde\beta_{ji})$ to $(0,\, \tilde\beta_{ji}^*)$.
\end{itemize}

In PCD-1 and PCD-2, we use the Newton-Raphson method to obtain $\tilde \beta_{ij}^*$ and $\tilde \beta_{ji}^*$. More generically, denoting the value of $\beta_{ij}$ at the $t^{\textrm{th}}$ iteration of PCD as $\beta_{ij}^{(t)}$, we update it to $\beta_{ij}^{(t+1)}=\beta_{ij}^{(t)}-\{\tilde V^{(1)}_j(\beta_{ij}^{(t)})+P_\lambda'(|\beta_{ij}^{(t)}|)\}/\{\tilde V^{(2)}_j(\beta_{ij}^{(t)})+P_\lambda''(|\beta_{ij}^{(t)}|)\}$, where $\tilde V^{(1)}_j(\beta_{ij}^{(t)})$ denotes $(\partial/\partial \beta_{ij}) \tilde V_j$ evaluated at $\beta_{ij}=\beta_{ij}^{(t)}$, $\tilde V^{(2)}_j(\beta_{ij}^{(t)})$ is equal to $(\partial^2/\partial \beta_{ij}^2) \tilde V_j$ evaluated at $\beta_{ij}=\beta_{ij}^{(t)}$, and $\beta_{ij}^{(0)}=\tilde\beta_{ij}$. Elaborating these derivatives gives the following formula leading to $\tilde\beta_{ij}^*$ at convergence, 
\begin{equation}
\beta_{ij}^{(t+1)}=\beta_{ij}^{(t)}-\frac{\bone^\T \bF_j (\bI_{n_{-j}}-\bQ_j)\bone+n_{-j}P_\lambda'(|\beta_{ij}^{(t)}|)}{\bone^\T\{(\bP_j-\bQ_j \bP_j-\bF_j \bF_j)(\bI_{n_{-j}}-\bQ_j)-\bF_j\bT_j\}\bone+0.5n_{-j}P_\lambda''(|\beta_{ij}^{(t)}|)}, \label{eq:NRupdate}
\end{equation}
where $\bone$ is an $n_{-j}\times 1$ vector of 1's, $\bI_{n_{-j}}$ is the $n_{-j}\times n_{-j}$ identity matrix, 
$\bF_j=\bC_j(\bC_j^\T \bC_j)^{-1}\bR_j^\T$, $\bP_j=\bR_j(\bC_j^\T \bC_j)^{-1}\bR_j^\T$, $\bQ_j=\bC_j(\bC_j^\T \bC_j)^{-1}\bC_j^\T$, $\bT_j=\bF_j+\bF_j^\T-\bQ_j \bF_j^\T-\bF_j \bQ_j$, in which \begin{equation*}
\bC_j=
\begin{bmatrix}
\bPsi_{j 1}^\T(\beta_{ij}^{(t)}) \\
\bPsi_{j 2}^\T(\beta_{ij}^{(t)}) \\
\vdots \\
\bPsi_{j, n_{-j}}^\T(\beta_{ij}^{(t)})
\end{bmatrix}, \, \, 
\bR_j=
\begin{bmatrix}
(\partial/\partial \beta_{ij}) \bPsi_{j 1}^\T(\tilde \bB_j)|_{\beta_{ij}=\beta_{ij}^{(t)}} \\
(\partial/\partial \beta_{ij}) \bPsi_{j 2}^\T(\tilde \bB_j)|_{\beta_{ij}=\beta_{ij}^{(t)}} \\
\vdots \\
(\partial/\partial \beta_{ij}) \bPsi_{j, n_{-j}}^\T(\tilde \bB_j)|_{\beta_{ij}=\beta_{ij}^{(t)}}
\end{bmatrix}, 
\end{equation*}
with $\bPsi_{j \ell}(\beta_{ij}^{(t)})$ denoting $\bPsi_{j\ell}(\tilde\bB_j)$ evaluated at $\beta_{ij}=\beta_{ij}^{(t)}$, 
$(\partial/\partial \beta_{ij}) \bPsi_{j \ell}(\tilde \bB_j)=-\bW[\ell, i]\bW^\T[\ell, -j]+\bSigma_u[-j, -j] \be_j$, and $\be_j$ is a $(p-1)\times 1$ vector whose entries are zero except for the entry corresponding to the location of $\beta_{ij}$ in $\bB_j$ being one. Finally, in (\ref{eq:NRupdate}), the first two derivatives of the SCAD penalty are 
\begin{align*}
P_\lambda'(|\beta_{ij}|)= & \left\{\lambda   I(|\beta_{ij}|\le \lambda)+ \frac{a\lambda-|\beta_{ij}|}{a-1}I( \lambda <|\beta_{ij}|\le a\lambda)\right\}\textrm{sgn}(\beta_{ij}),\\
P_\lambda''(|\beta_{ij}|)= & -\frac{1}{a-1}I( \lambda <|\beta_{ij}|\le a\lambda),
\end{align*}
where $\textrm{sgn}(t)=I(t>0)-I(t<0)$. 

In PCD-3 and PCD-4, we choose between $(\tilde \beta_{ij}^*,\, 0)$ and $(0,\, \tilde \beta_{ji}^*)$ to decide if $X_i$ is a parent of $X_j$ or the other way around. The choice is made based on the sum of the two (partially updated) penalized objective functions, one associated with $X_j$ and the other associated with $X_i$, evaluated at each pair. The pair leading to a smaller sum is chosen as the updated value of $(\tilde \beta_{ij}, \, \tilde\beta_{ji})$.  If the chosen pair contains a nonzero component smaller than a pre-specified threshold in absolute value, such as $10^{-4}$, we conclude that there is no edge between the two nodes. 

\setcounter{equation}{0}		
\section*{Appendix C: The PCD algorithm for the naive likelihood-based method}
\setcounter{section}{0}
\renewcommand{\theequation}{C.\arabic{equation}}

One can follow a similar algorithm described in Appendix B to find the miminizer of the naive penalized log-likelihood function $R_{\textrm{nv}}(\bB)$ in equation (5) in the main article that relates to a DAG. In this case, one may use the naive least square estimate of $\bB_j$, for $j=1, \ldots, p$, to construct an initial estimate for $\bB$ in Step 1. In Step 2, one would replace $\tilde V_j$ and $\tilde V_i$ above by $\tilde V_{j, \textrm{nv}}$ and $\tilde V_{i, \textrm{nv}}$, respectively, where  
\begin{equation*}
\tilde V_{j,\textrm{nv}}=\frac{n_{-j}}{2}\log\left\{\sum_{\ell\in O_j}\left(\bW[\ell, j]- \sum_{k\ne i,j} \bW[\ell, k] \tilde \beta_{kj} -\bW[\ell, i] \beta_{ij}\right)^2 \right\},
\end{equation*}
and $\tilde V_{i, \textrm{nv}}$ is similarly defined. After these replacements, $\tilde\beta^*_{ij}$ and $\tilde \beta^*_{ji}$ in PCD-1 and PCD-2 are solutions to polynomial equations of order $r$, where $r\le 3$. Hence, $\tilde\beta^*_{ij}$ and $\tilde\beta^*_{ji}$ can be found explicitly (when $r<3$) or computed numerically via a polynomial equation solver (when $r=3$). In particular, the equation to solve in order to find 
\begin{equation*}
\tilde\beta_{ij}^*=\argmin_{\beta_{ij}}\left\{\tilde V_{j,\textrm{nv}}+\sum_{k\ne i,j}P_{\lambda}(|\tilde\beta_{kj}|)+P_\lambda(|\beta_{ij}|)\right\}
\end{equation*}
is given by 
\begin{eqnarray}
0 & =& n_{-j}\sum_{\ell \in O_j} \left(\bW[\ell, j]-\sum_{k\ne i,j}\bW[\ell, k]\tilde\beta_{kj}-\bW[\ell, i]\beta_{ij}\right) \bW[\ell, i]-\nonumber \\
  & & P_\lambda'(|\beta_{ij}|)\textrm{sgn}(\beta_{ij})\sum_{\ell \in O_j}\left(\bW[\ell, j]-\sum_{k\ne i,j}\bW[\ell, k]\tilde\beta_{kj}-\bW[\ell, i]\beta_{ij}\right)^2. 
\label{eq:raweq}
\end{eqnarray}

If a solution exists that satisfies $|\tilde \beta_{ij}^*|\ge a\lambda$, then (\ref{eq:raweq}) reduces to a linear equation in $\beta_{ij}$, and the solution can be trivially found to be 
\begin{equation*}
\tilde\beta_{ij}^*=\frac{\sum_{\ell \in O_j}(\bW_[\ell, j]-\sum_{k\ne i,j} \bW[\ell, k]\tilde\beta_{kj}) \bW[\ell, i]}{\sum_{\ell \in O_j} \bW[\ell, i]^2}.
\end{equation*}

If a non-zero solution exists in $[-\lambda, \, \lambda]$, (\ref{eq:raweq}) is a quadratic equation, $c_2 \beta_{ij}^2+c_1\beta_{ij}+c_0=0$, and  $\tilde\beta_{ij}^*$ is the root given by $(-c_1+\sqrt{c_1^2-4 c_2 c_0})/(2 c_2)$, where 
\begin{align*}
c_2 & = \textrm{sgn}(\tilde\beta_{ij}^*) \lambda \sum_{\ell \in O_j}\bW[\ell, i]^2, \\
c_1 & = 2c_2+n_{-j}\sum_{\ell \in O_j} \bW[\ell, i]^2-\textrm{sgn}(\tilde \beta_{ij}^*) 2 \lambda \sum_{\ell \in O_j} \bW[\ell, i] \sum_{k \ne i,j} \bW[\ell, k] \tilde \beta_{kj},\\
c_0 & = \textrm{sgn}(\tilde\beta_{ij}^*)\lambda \sum_{\ell \in O_j}\left(\bW[\ell, j]-\sum_{k\ne i, j}\bW[\ell, k]\tilde \beta_{kj}\right)^2-n_{-j}\sum_{\ell \in O_j}\left(\bW[\ell, j]-\sum_{k \ne i, j} \bW[\ell, k]\tilde \beta_{kj}\right)\bW[\ell,i].
\end{align*}
At the first glance, it may seem strange to have $c_2$, $c_1$, and $c_0$ depend on the solution via $\textrm{sgn}(\tilde \beta_{ij}^*)$. It actually  is a way to check the existence of a minimizer of the optimization problem. It is only when the sign of the root $(-c_1+\sqrt{c_1^2-4 c_2 c_0})/(2 c_2)$ agrees with the value of $\textrm{sgn}(\tilde\beta_{ij}^*)$ used in $c_2$, $c_1$, and $c_0$ do we claim that a non-zero minimizer in $[-\lambda, \, \lambda]$ is found, and it is equal to this root.

Lastly, if a solution exists that satisfies $\lambda<|\tilde\beta_{ij}^*|\le a\lambda$, then (\ref{eq:raweq}) is a cubic equation, $c_3\beta_{ij}^3+c_2 \beta_{ij}^2+c_1\beta_{ij}+c_0=0$, where 
\begin{align*}
c_3 & = \frac{1}{a-1}\sum_{\ell \in O_j} \bW[\ell, i]^2, \\
c_2 & = -\textrm{sgn}(\tilde\beta_{ij}^*)a \lambda c_3-\frac{2}{a-1}\sum_{\ell \in O_j}\left(\bW[\ell, j]-\sum_{k \ne i, j} \bW[\ell, k]\tilde \beta_{kj}\right)\bW[\ell,i], \\
c_1 & = -n_{-j}(a-1) c_3-a \lambda \left\{a\lambda c_3 +\textrm{sgn}(\tilde\beta_{ij}^*) c_2\right\}+\frac{1}{a-1}\sum_{\ell \in O_j}\left(\bW[\ell, j]-\sum_{k \ne i, j} \bW[\ell, k]\tilde \beta_{kj}\right)^2, \\
c_ 0 & = -\frac{a-1}{2}n_{-j}\left\{\textrm{sgn}(\tilde\beta_{ij}^*)a\lambda c_3+c_2\right\}-\textrm{sgn}(\tilde \beta_{ij}^*)\frac{a\lambda}{a-1}\sum_{\ell \in O_j}\left(\bW[\ell, j]-\sum_{k \ne i, j} \bW[\ell, k]\tilde \beta_{kj}\right)^2.
\end{align*}
The solution is the real root of this equation. 
		
\setcounter{equation}{0}		
\section*{Appendix D: Proof of the consistency of SIC}
\setcounter{section}{0}
\renewcommand{\theequation}{D.\arabic{equation}}

Recall that the tuning parameter selector is, with the subscript $n$ added to highlight its dependence on $n=\min_{1\le j \le p}n_{-j}$,
\begin{equation}
\textrm{SIC}_n(G)=\sum_{j=1}^p\left(\hat V_j+e_j\frac{\log n_{-j}}{n_{-j}}\right), \label{eq:SICG}
\end{equation}
where $e_j$ is the number of parents of $X_j$ according to $G$, and $\hat V_j$ is equal to $V_j$ evaluated at the unpenalized corrected score estimate of $\bB_j$ given the structure of $G$. For notational simplicity, we assume $n_{-j}=n$, for $j=1, \ldots, p$, in the sequel, and the proposed information criterion can be re-expressed as
$$\textrm{SIC}_n(G) = \sum_{j=1}^p \hat V_j+e_{G}(\log n)/n  = Q_n(G)+P_n(G),$$
where $e_{G}$ is the number of edges in $G$, $Q_n(G)=\sum_{j=1}^p \hat V_j$ is the sum of $p$ quadratic forms that depends on the unpenalized estimate of $\bB$ given the structure of $G$, and $P_n(G)=e_{G}(\log n)/n$ assesses the complexity of $G$ while accounting for the sample size. Denote by $\bE_{G}$ the set of directed edges in $G$, and by $|\bE_{G}|$ the size of this set, i.e., $|\bE_{G}|=e_{G}$.

To prove the consistency of $\textrm{SIC}_n(G)$, it suffices to establish the following two assertions,
\begin{eqnarray}
\textrm{SIC}_n(G_-)-\textrm{SIC}_n(G_0) & > 0 & \textrm{ with probability approaching one, as $n\to \infty$}, \label{eq:underfit}  \\
\textrm{SIC}_n(G_+)-\textrm{SIC}_n(G_0) & \to 0^+ & \textrm{ in probability as $n\to \infty$.} \label{eq:overfitm} 
\end{eqnarray}
If one allows $p$ to increase as $n\to \infty$, in order to show (\ref{eq:underfit}) and (\ref{eq:overfit}), we also need to assume (A6) $p^2 (\log n)/n \to 0$ as $n\to \infty$, besides regularity conditions (A1)--(A3) listed in Web Appendix A. Without imposing sparsity assumption on $G$, $e_{G}$ is at most $p(p-1)/2=O(p^2)$, and (A6) guarantees the penalty $P_n(G)$ for graph complexity shrinks to zero as $n\to \infty$ even for dense graphs. In addition, given (A2), (A6) is a sufficient condition for $Q_n(G)$ to be bounded. In this study we assume $e_{G_0}<p(p-1)/2$. 

To show (\ref{eq:underfit}), we will relate graph selection with variable selection in a regression model given a set of candidate predictors as follows. Denote by $\mathscr{M}_j$ a regression model with $X_j$ being the response variable and the remaining nodes as potential predictors. One selecting an underfitted graph, $G_-$, means that, for at least one edge in $G_0$, one either reverses it or completely misses it in the selected graph. Suppose in $G_0$, there is an edge pointing from $X_i$ to $X_j$, where $i \ne j$, and it is reversed in the selected $G_-$; this means that one underfits the regression model $\mathscr{M}_j$ and overfits the regression model $\mathscr{M}_i$. If this directed edge in $G_0$ is missing in $G_-$, it means that one underfits $\mathscr{M}_j$. In conclusion, whenever one selects an underfitted graph, one must have underfitted $\mathscr{M}_j$ for at least one $j\in \{1, \ldots, p\}$. For such $\mathscr{M}_j$, under (A1)--(A2), the $j$th summand in $Q_n(G_-)$ is strictly positive in probability as $n\to \infty$, whereas each summand in $Q_n(G_0)$ converges to zero in probability. It follows that, in 
\begin{eqnarray*}
\textrm{SIC}_n(G_-)-\textrm{SIC}_n(G_0)& = & \{Q_n(G_-)-Q_n(G_0)\}+\{P_n(G_-)-P_n(G_0)\} \\
& = & \{Q_n(G_-)-Q_n(G_0)\}+(e_{G_-}-e_{G_0})(\log n)/n,
\end{eqnarray*}  
the first difference is positive in probability, which is bounded even when $p\to \infty$ given (A6), and the second difference converges to zero as $n \to \infty$ under (A6). Hence (\ref{eq:underfit}) holds. 

To show (\ref{eq:overfit}), note that, one selecting an overfitted graph, $G_+$, is equivalent to one choosing an overfitted model for $\mathscr{M}_j$ for at least one node $X_j$. Under (A1)--(A3), overfitting $\mathscr{M}_j$ does not inflate the $j$th summand in $Q_n(G_+)$ in probability, which is the key difference from underfitting $\mathscr{M}_j$ considered earlier. More importantly, by \citet{Hansen82}, evaluated at the overfitted $\mathscr{M}_j$ or the true $\mathscr{M}_j$ both yield $\hat V_j$ converging to a quantity of order $O_{P}(n^{-1})+o_{P}(\log n/n)$, as $n\to \infty$. Hence, in
\begin{eqnarray*}
\textrm{SIC}_n(G_+)-\textrm{SIC}_n(G_0)& = & \{Q_n(G_+)-Q_n(G_0)\}+\{P_n(G_+)-P_n(G_0)\} \\
& = & \{O_{P}(p/n)+o_{P}(p\log n/n)\}+(e_{G_+}-e_{G_0})(\log n)/n,
\end{eqnarray*}  
given (A6), the first two terms in conjunction converge to zero in probability, and so does the latter difference, and latter difference tends to zero from above since it is strictly positive for all $n$ and $p$. Hence (\ref{eq:overfit}) holds. This completes the proof that $\textrm{SIC}_n(G)$ is a consistent information criterion for selecting DAGs. The arguments in this appendix still carry over following similar ideas when $\{n_{-j}\}_{j=1}^p$ are not all the same and $n=\min_{1\le j  \le p}n_{-j}$.

\bibliographystyle{apalike}

\bibliography{references}

\end{document}